\def\micron{{$\mu$}m}
\def\BJ{B$_{\rm J}$}

\def\to{\hbox{$\,$--$\,$}}
\def\muspc{\hskip 0.15 em}

\def\mag{\hbox{$\;.\!\!\!^m$}}
\input psfig
\documentclass{aapress}
\begin{document}

\thesaurus{06(19.34.1; 19.44.1; 19.67.1)}

\title{Field \#3 of the Palomar-Groningen 
survey\thanks{Based on observations obtained at the European Southern
Observatory, La Silla, Chile}}
\subtitle{II. Near-infrared photometry of semiregular 
variables\thanks{Table 2 is only available in electronic form at the CDS 
via anonymous {\tt ftp} to {\tt cdsarc.u-strasbg.fr (130.79.128.5)}
or {\tt WWW} at {\tt URL} http://cdsweb.u-strasbg.fr/Abstract.html
}}

\author{M. Schultheis\inst{1,3} 
  \and Y.K. Ng\inst{2,3,4}\thanks{{\em Present address:}\/
{Space Research Organisation Netherlands
(SRON), Sorbonnelaan 2, 3584~CA \ Utrecht, the Netherlands}}
  \and J. Hron\inst{1}
  \and F. Kerschbaum\inst{1}
}

\institute{Institut f\"ur Astronomie der Universit\"at Wien,
           T\"urkenschanzstra{\ss}e 17, A-1180 Wien, Austria
  \and Leiden Observatory, P.O. Box 9513, 2300 RA \ Leiden, the Netherlands
  \and Institut d'Astrophysique de Paris, CNRS,
           98bis Boulevard Arago, 75014 Paris, France 
  \and Padova Astronomical Observatory,
           Vicolo dell'Osservatorio 5, I-35122 Padua, Italy \\
\null ({\tt e-mail: (hron,kerschbaum,schultheis)\char64astro.ast.univie.ac.at;
schulthe\char64iap.fr; yuen\char64pd.astro.it})
}

\offprints{M. Schultheis}
\mail{schulthe@iap.fr}

\date{Received 10 May 1996 / Accepted 22 July 1998} 

\maketitle   

\markboth{M. Schultheis et al.:\ Field \#3 of the Palomar-Groningen Survey II}
{Near-infrared photometry of semiregular variables}

\begin{abstract} 
Near-infrared photometry (JHKL$^\prime$M) was obtained for
78 semiregular variables (SRVs) in field \#3 of the Palomar-Groningen
survey (PG3, $l\!=\!0\degr$, $b\!=\!-10\degr$). 
Together with a sample of Miras in this field a comparison 
is made with a sample of field SRVs and Miras.
The PG3 SRVs form a sequence (period\to{luminosity} \& period\to{colour})
with the PG3 Miras, 
in which the SRVs are the short period extension to the Miras.
The field and PG3 Miras follow the same P/(J--K)$_0$ relation,
while this is not the case for the field and PG3 SRVs.
Both the PG3 SRVs and Miras follow the Sgr~I period-luminosity relation
adopted from Glass et~al.\ (\cite{Glass95ea}). They are likely
pulsating in the fundamental mode 
and have metallicities spanning the range 
from intermediate to approximately solar. 

\end{abstract}

\hyphenation{Kersch-baum}
\hyphenation{Miras}
\section{Introduction}
\label{Introduction}
In the last decade much effort has been spent on the studies of Miras
(e.g. van der Veen \cite{vdVeen88}, Feast et~al.\ \cite{Feast89ea}, 
Whitelock et~al.\ \cite{Whitelock91ea}, Blommaert \cite{Blommaert92}, and
Glass et~al.\ \cite{Glass95ea}), OH/IR stars 
(e.g. Herman \cite{Herman88},
te Lintel-Hekkert \cite{Lintel90}, 
Lindquist et~al.\ \cite{Lindquist92ea},
van Langevelde \cite{vLangevelde92}, 
Blommaert et~al.\ \cite{Blommaert94ea}, and
Habing \cite{Habing96}), 
and carbon stars 
(e.g. Willems \cite{Willems87}, 
Chan\mbox{\muspc\&\muspc}Kwok \cite{ChanKwok88}, 
Willems \cite{Willems88a}b, 
Willems\mbox{\muspc\&\muspc}de~Jong \cite{WillemsdeJong88}, 
de~Jong \cite{deJong89}, 
Stephenson \cite{Stephenson89}, 
Zuckerman\mbox{\muspc\&\muspc}Maddalena \cite{ZuckMad89},
Azzopardi et~al.\ \cite{Azzopardi91ea}, 
Tyson\mbox{\muspc\&\muspc}Rich \cite{TR91}, 
Westerlund et~al.\ \cite{Westerlund91ea}, 
Groenewegen et~al.\ \cite{Groenewegen92ea}, 
Groenewegen \cite{Groenewegen93}, Marigo et~al.\ \cite{Marigo96bea}, 
Ng \cite{Ng97b} \& \cite{Ng98}, and
Marigo \cite{Marigo98}).
\par 
Systematic investigations of 
semiregular variables (SRVs) 
were carried out by Kerschbaum\mbox{\muspc\&\muspc}Hron (\cite{KH92}, 
\cite{KH94};
hereafter respectively referred to as KH92\mbox{\muspc\&\muspc}KH94)
and Jura\mbox{\muspc\&\muspc}Kleinmann (\cite{JuraKlein92}) 
in recent years. 
From the temperatures, luminosities, mass loss rates and
number densities KH92\mbox{\muspc\&\muspc}KH94
distinguished a `blue', `red', and a `Mira-like'
group among their SRVs. 
Their result suggests an evolutionary
sequence, where the `blue' SRVs evolve towards the thermally
pulsing AGB, change into `red' SRVs,   
and then continue to evolve to the Mira phase. 
\par
AGB stars are ideal probes in studies of the galactic bulge, 
because of their high luminosities. 
Various photometric
(Frogel\mbox{\muspc\&\muspc}Whitford \cite{FW87}, 
Terndrup \cite{Terndrup88}, 
Geisler\mbox{\muspc\&\muspc}Friel \cite{GeislerFriel92}), 
spectroscopic (Rich \cite{Rich90}, 
McWilliam\mbox{\muspc\&\muspc}Rich \cite{WR94}, 
Sadler et~al.\ \cite{Sadler96ea})
and star counts (Ng et~al.\ \cite{Ng96aea})
studies from Baade's Window (hereafter referred to 
as BW; $l\!=\!1\fdg0$, $b\!=\!-3\fdg9$),
show a large spread in metallicity ranging from 
--1.5\muspc$<$\muspc{Fe/H}\muspc$<$\muspc1.0.
The bulge offers a good opportunity to study the evolutionary
aspects of AGB-stars in relation with their metallicity,
which is one of the major evolutionary parameters.
\par
\begin{figure}
\vbox{\null\vskip6.8cm
\includegraphics{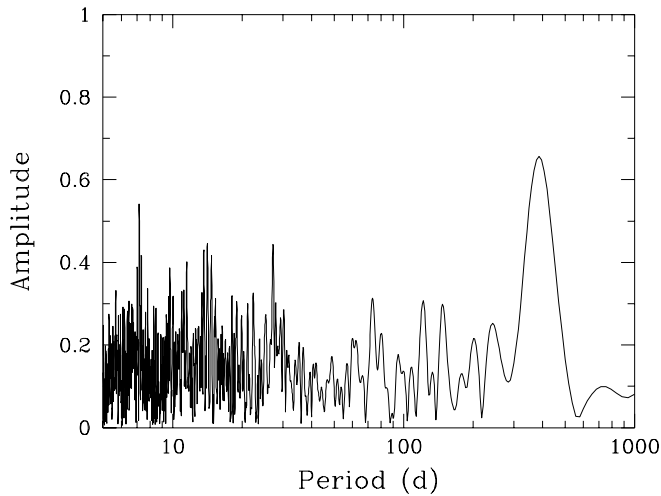}
}
\caption{The Fourier spectral window for the \BJ\ plate coverage 
of PG3 (Wesselink \cite{Wesselink87})}
\label{fig1}
\end{figure}
PG3 (field \#3 of the Palomar-Groningen Variable Star Survey; 
$l\!=\!0\degr$, $b\!=\!-10\degr$; see Sect.~\ref{PG3} for details) 
is well searched for variable stars (Plaut \cite{Plaut71}, 
Wesselink \cite{Wesselink87}; 
the latter will hereafter be referred to as Wess87). 
Blommaert (\cite{Blommaert92}; hereafter referred to as Bl92) 
studied the properties of the PG3 Miras
with nIR (near-infrared) 
and IRAS photometry. The period-colour relation for the 
PG3 Miras was found to be shifted significantly 
from the Large Magellanic Cloud (LMC) relation 
(Feast et~al.\ \cite{Feast89ea}). 
With the theoretical relation from Wood et~al.\ (\cite{Wood91ea}) 
it was concluded that this must be due to 
metallicity, which could be about 1.6 times larger for 
the PG3 Miras (Bl92). 
Feast (\cite{Feast96}) obtained similar results for the Miras in 
the galactic bulge versus those in the LMC. 
Comparable indications were found by
Whitelock (\cite{Whitelock96}) from the local Miras.
\par
In this study a comparison is made 
between the pulsation and near-infrared 
properties of the PG3 SRVs and Miras and the field SRVs and Miras. 
In Sect.~\ref{Data} a discussion is given about the SRV data: 
the observations and the 
comparison samples. We proceed in Sect.~\ref{Results} with a 
description of the results of this analysis, continue 
with the discussion in Sect.~\ref{Discussion} and summarize it 
in Sect.~\ref{Summary}.
\par

\section{Data}
\label{Data}
\subsection{PG3}
\label{PG3}
Four fields were selected in the mid-fifties by Baade and Plaut to
search for variable stars (Blaauw \cite{Blaauw55}; 
Larsson-Leander \cite{LarssonLeander59}).
The results from the photographic survey, known as the 
Palomar-Groningen Variable Star survey, were published in 
a series of six papers (Plaut \cite{Plaut66}, 
\cite{Plaut68a}b, \cite{Plaut70}, \cite{Plaut71}, 
\cite{Plaut73}). 
The centre of PG3 is located 10$^\circ$ south of the galactic centre
and skims over the edge of the galactic bulge.
Important aspects of this field are the large area covered
($6\fdg5\times6\fdg5$) and the
low interstellar extinction, which is gradually increasing 
in the direction towards 
the galactic centre (Wess87).
\par
With emphasis on the RR Lyrae stars, the variable stars in PG3
were re-examined by Wess87, using 
UKST B$_{\rm J}$ and R$_{\rm F}$ Schmidt plates.
Figure~\ref{fig1} shows the Fourier spectral window for the 
epochs of the \BJ\ plates. The figure shows
that a good resolution, as intended, is obtained towards the short 
period variables, but for the long period variables
a deficiency of stars with
periods between 320 to 500 days could be possible. The 
three highest alias peaks correspond in decreasing order
with one year, one week, and one synodical month.
\par
\begin{table}
\caption{Log of the observing runs}
\begin{tabular}{|cl|ccll|}
\hline
Run & & dates & year & observer(s) & \\
\hline
1 & & 28 Jul -- 2 Aug& 1990 & Ng &\\
2 & & 21 May -- 3 Jun& 1991 & Brown &\\
3 & & 20 -- 29 Aug& 1991 & Ng &\\
4 & & 16 -- 21 Jun& 1992 & Ng \& Schultheis &\\
5 & & 29 Jun -- 3 Jul& 1993 & Ng \& Schultheis &\\
\hline
\end{tabular}
\end{table}
The large number of Miras and SRVs discovered
in this field makes it very attractive, to subject these stars to a
more detailed study. 
Bl92 studied a sample
of Miras and compared them with the IRAS sources in PG3,
while we focus on the SRVs.

\subsection{The SRV sample}
In the GCVS4 catalog (Kholopov et~al.\ \cite{GCVS4}) 
the classification of SRVs
is based on the shape and the amplitude of the light curve. 
Generally, the period of the variations ranges from 20 to 2000 days
with an amplitude less than \mbox{V\muspc=\muspc2\mag5}. 
Plaut (\cite{Plaut71}) distinguished in his classification SRa and SRb 
type variables, while Wess87 made no distinction.
Wess87 based his criteria on the 
B$_{\rm J}$ and R$_{\rm F}$ amplitudes, smaller 
than 2\mag0, and periods ranging from \mbox{$\sim$\muspc30\to1000} days. 
The SRV classification of Wess87 is in most of the cases
compatible with a SRV of type `a' 
(hereafter referred to as SRaV)
from Plaut (\cite{Plaut71}). 
In general, the lightcurves of the SRaVs resemble
those from the Miras. 
The difference in the classification 
is merely a consequence of the imposed amplitude limits in the variation. 
\par
In this study the SRV stars are selected with the 
Wess87 classification.
Higher priority is given to the observations of stars with 
a Q\muspc=\muspc0 quality flag, indicating that there is no doubt about
the classification and the period.
The SRVs are selected such, that there is no bias to  
the brightest SRVs and that
the sample covers in 25 days intervals 
the full period range. 
\par
\begin{center}
\begin{table*}
\caption{Near-infrared photometry for
the stars in field \#3 of the Palomar-Groningen Variable Star
Survey (Plaut 1971). Column 1 lists the stellar identifier, 
adopted from Wesselink (1987);
column 2 gives the identification made by Plaut (1971);
column 3--7 gives the JHKL$^\prime$M photometry,
typical errors are $\sim$ 0\mag02 in JHK, $\sim$ 0\mag1 in L$^\prime$
and $>$ 0\mag2 in M;
column 8 gives the observing run identifier (see table 1);
column 9 gives the period determined by Wesselink (1987)
if available; and column 10 gives the quality flag related 
with the period and the identification of the star 
(Q=0: no doubt about the determined period and classification,
Q=1: classification is correct but alternative period is possible,
Q=2: period determination is correct but the classification is doubtful,
Q=3: both period determination and classification are unreliable)}
\begin{center}
{The table is available via anonymous {\tt ftp} to 
{\tt cdsarc.u-strasbg.fr (130.79.128.5)}}
\end{center}
\end{table*}

\end{center}

\subsection{Near-infrared Photometry}
\label{NIRphot}
Near-infrared photometry (JHKL$^\prime$M) of 78 PG3 SRVs 
was obtained with the ESO 1-m telescope, La Silla (Chile),
equipped with an InSb detector. 
The observations were carried out 
under photometric circumstances in the observing seasons
1990\to1993 (ESO N$^o$ 49.5\to011 \& 51.7\to056). 
Table~1 shows the log of the observing runs. 
The 1990\to1991 observations (see Bl92) of the stars 
S40, S147, S728, S969, S1008, S1016, S1128 and S1204
were carried out 
as part of the ESO key programme `Stellar evolution in the
galactic bulge' (Blommaert et~al.\ \cite{Blommaert90ea}; 
ESO N$^o$ 45K.5\to007).
The observations were made in a standard way,
through a diaphragm with a 15$\arcsec$ aperture
with the chopping and beam-switching
technique.
An 8\,Hz sky chopping and a beam-switch throw of approximately
20$\arcsec$ in R.A. was applied.
\par
The JHKL$^\prime$M fluxes are calibrated to the ESO photometric system
(Bouchet et~al.\ \cite{Bouchet91ea}).
The typical errors are
$\sim$\muspc0\mag02 in JHK, $\sim$\muspc0\mag1 in$\rm ^{\null}$  L$^\prime$,
and $>$\,0\mag2 in M.
For some stars no L$^\prime$ and M
photometry was obtained, because they were fainter than the limiting
magnitude attainable with the telescope.
Table~2 gives the JHKL$^\prime$M magnitudes of the PG3 SRVs
together with the period (Wess87).
\par
In general the limiting magnitude was 
K$_{lim}$\muspc$\simeq$\muspc10$^m$ for most of our observing
nights, which is due to our relatively
short integration time and integration sequence.
Photometry for some of the fainter stars were obtained 
in nights when the photometric conditions allowed it.
In this respect the DENIS survey (Epchtein et~al.\ \cite{Epchtein94ea},
\cite{Epchtein97ea}),
with \mbox{K$_{lim}$\muspc$\simeq$\muspc14$^m$}, should be able to
improve on the photometry and reach significantly deeper limits.

\subsection{Crowding}
\label{Crowding}
The presence of additional stars in the beam cannot be avoided,
because we are dealing with crowded field observations.
Two cases should be considered: additional stars in the 
primary or in the background beam.
An additional flux contribution in the primary beam would lead to
an increase of the flux from the star, 
while stars in the background beams 
would give a background subtraction which is too high. 
As a consequence the flux of the target star will be underestimated.
In first approximation, the number of faint stars  
in both the primary beam and the background beams are similar.
\par
The stars in the background are in general much fainter and not as red as 
the stars observed. The induced errors from stars present in 
the background beams
are expected to be less than the errors quoted above. 
A new position for the background subtraction would have been selected, 
if a bright star was noticed in one of the background beams,
but this was not necessary. 
The flux of the target star will be overestimated with 
additional stars in the primary beam. 
Due to stars surrounding 
S283 (see finding chart in Ng\mbox{\muspc\&\muspc}Schultheis \cite{NS97}), 
the magnitudes are slightly too bright and the colours
too blue for this star.
\par

\subsection{Interstellar extinction}
\label{extinction}
The procedure described by Bl92 is used,
to correct for the interstellar extinction.
It is based on the PG3 extinction map, constructed by Wess87
from the colour excess of the RR Lyrae stars at minimum
light. In this map the extinction is highest
in the plate corner at lowest galactic latitude
(A(\BJ)\muspc=\muspc1\mag14; $b\!=\!-6\degr$\/)
and lowest at the opposite corner
(A(\BJ)\muspc=\muspc0\mag14; $b\!=\!-14\degr$\/).
The extinction is
described with a linear relation, because of the smoothness
of the gradient in the extinction map. In first approximation we have
A(\BJ)\muspc=\muspc$0.133\,b + 2.02$, where $b$\/ is the
galactic latitude.
For a normal extinction law the
corrections for the different infrared passbands
can be derived with the standard curve no.~15 of
van~de~Hulst (\cite{vdHulst49}). All the JHK photometry
for the PG3 stars discussed in this study are de-reddened
with the procedure outlined above.
The photometric data in Table~2 are not de-reddened. 
\par

\subsection{The comparison sample}
\label{comparison}
A comparison is made with 
other near-infrared photometric studies of SRVs 
and Miras, in order to obtain a better understanding of the evolutionary
status of the PG3 SRVs. 
We use a sample of PG3 Miras (Bl92), well observed
O-rich field Miras (Catchpole et~al.\ \cite{Catchpole79ea})
and a 
magnitude limited sample of field O-rich SRVs (KH94, 
Kerschbaum \cite{Kerschbaum95}). 
Note, that the field samples mentioned above
might not be truly representative in their relative numbers 
for the local neighbourhood.
The field stars were de-reddened with a procedure similar to 
Feast et~al.\ (\cite{Feast82ea}). 
The reddening corrections are small, because
the visual absorptions ranges typically 
from 0\mag05\to0\mag20.
We further used the Sgr~I Miras from Glass et~al.\ (\cite{Glass95ea}) and 
the LMC
LPVs (long period variables)
from Reid et~al.\ (\cite{Reid95ea})
as an extra indication. 
For the latter 
we defined Mira\mbox{\muspc\&\muspc}SRV variable groups, following 
Hughes\mbox{\muspc\&\muspc}Wood (\cite{HW90})
on basis of the I-band amplitudes.
We use the Reid et~al.\  data set instead 
of the Hughes\mbox{\muspc\&\muspc}Wood data set, because 
the former are in the same photometric system as the Sgr~I Miras.
\par
\begin{figure}
\vbox{\null\vskip6.5cm
\includegraphics{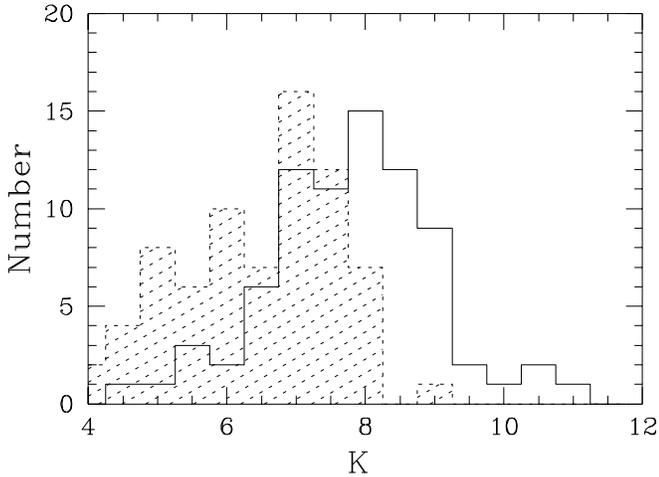}
}
\caption{Distribution of K-magnitudes for the SRVs (solid histogram; 
this paper) 
and the Miras (dot-shaded histogram; Bl92) observed in PG3}
\label{fig2}
\end{figure}

\subsection{Photometric transformations}
\label{transformation}
The photometry of the PG3 SRVs\mbox{\muspc\&\muspc}Miras and the field SRVs 
are in the ESO photometric system 
(Bouchet et~al.\ \cite{Bouchet91ea}, 
van der Bliek et~al.\ \cite{vdBliek96ea}).
The photometry for the field Miras was obtained in the SAAO
photometric system as defined by Glass (\cite{Glass74}).
This photometric system is not identical to the 
SAAO system in which the photometry for the 
Sgr~I Miras, the LMC LPVs, and
the period-luminosity \& period-colour relations were obtained 
(Glass et~al.\ \cite{Glass95ea}).
All the photometry discussed in this paper are in the ESO photometric system
(Bouchet et~al.\ \cite{Bouchet91ea})
or transformed to it from the various SAAO systems. 
New transformations from Hron et~al.\ (\cite{Hron98ea}) are used,
because existing transformations either refer to the old
ESO system 
(Bessell\mbox{\muspc\&\muspc}Brett 1988, Carter \cite{Carter90}, 
Engels et~al.\ \cite{Engels81ea}, Wamsteker \cite{Wamsteker81})
or do not cover all SAAO systems
(Bessell\mbox{\muspc\&\muspc}Brett \cite{Bessell88}, 
Bouchet et~al.\ \cite{Bouchet91ea}, 
Carter \cite{Carter90},
van der Bliek et~al.\ \cite{vdBliek96ea}).
Furthermore, these transformations
do not include stars with \mbox{(J--K)\muspc$>$\muspc1\mag0}.
Extrapolation of these transformations to the colours of 
typical AGB stars leads to errors 
of the same order as e.g.\ the differences due to metallicity
(Hron et~al.\ \cite{Hron98ea}).
The estimated uncertainties in the new transformations 
are typically \mbox{$\sim$\muspc0\mag02} in the colours.
\par
%
\begin{figure}
\vbox{\null\vskip8.8cm
\includegraphics{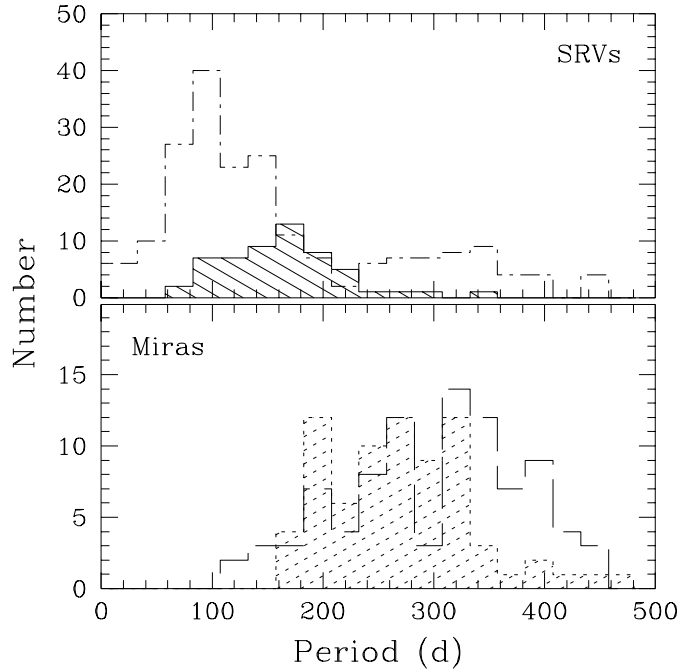}
}
\caption{Period distribution for the SRVs (solid, shaded histogram; this paper)
and the Miras (dot-shaded histogram; Bl92) 
observed in PG3 together with the distributions for
the field SRVs (dot-dashed histogram; KH94 and
Kerschbaum \cite{Kerschbaum95}) 
and Miras (long dashed histogram; Catchpole et~al.\ \cite{Catchpole79ea})}
\label{fig3}
\end{figure}

\section{Results}
\label{Results}
\begin{figure}
\vbox{\null\vskip6.5cm
\includegraphics{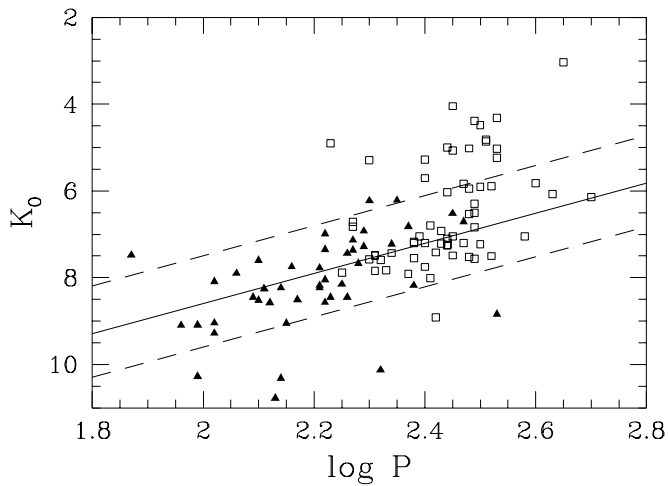}
}
\caption{K$_0$ vs log P relation for the PG3 SRVs 
(solid triangles; this paper)
and PG3 Miras (open boxes; Bl92).
The straight line shows the Sgr~I relation given by 
Glass et~al.\ (\cite{Glass95ea}).
The two dashed lines indicate the transition between
bulge and non-bulge stars (see Sect.~\ref{PKrelation} for details) }
\label{fig4}
\end{figure}

The observations are presented in Figs.~\ref{fig2}\to\ref{fig8}. 
The mean magnitude is used in the figures,
when several observations are available for a star.
In Figs.~\ref{fig4}\to\ref{fig8} only those stars are considered which, 
according
to Wess87, are classified correctly and have a good period determined.
\par

\subsection{Magnitude Distribution}
\label{magdistribution}
At present we are mainly interested
in the bulge stars. Therefore, disc stars are referred to
as the foreground contamination. 
If the stars were located in the disc, the magnitude distribution 
of the foreground contamination 
in PG3 will show a gradually, more or less linear, 
increase towards fainter magnitude (see for example 
Fig.~11\to4, Ng et~al.\ \cite{Ng95ea}).
The roughly linear increase is a consequence of the increasing 
volume in the cone, when sampled
towards larger distances. 
The distribution of stars in the bulge 
have in first approximation
a peaked shape (see for example Fig.~11\to3, Ng et~al.\ \cite{Ng95ea}).
This is a result of the density profile 
of the bulge stars, which increases towards 
the galactic centre, is highest near the galactic centre,
and decreases afterwards.
\par
Two interpretations are possible for the K-magnitude
distributions displayed in Fig.~\ref{fig2}.
\par
\noindent
1.\quad
There is a difference in the distribution
of the foreground contamination (K\muspc$<$\muspc6\mag0), 
between the SRVs and Miras.
The foreground contamination of the Miras appears to be significantly
larger than for the SRVs. The large fraction of foreground 
Miras was already noticed by Bl92, especially 
those which also have an IRAS 12 \micron\ and \mbox{25~\micron\
detection}. Moreover, Fig.~\ref{fig2}  shows that the peak of the
SRVs (6\mag5\muspc$<$\muspc{K}\muspc$<$9\mag0) is about 1.5
times broader than the Mira peak between
6\mag5\muspc$<$\muspc{K}\muspc$<$8\mag0.
This could indicate that part of the SRVs 
with 6\mag5\muspc$<$\muspc{K}\muspc$<$8\mag0 are in
fact Mira type variables. They were classified as SRVs, 
because of the amplitude of the variations in the lightcurves.
In this case some of the SRVs with K\muspc$>$\muspc8\mag0 represent
a group of intrinsically fainter stars. 
\par
\noindent
2.\quad
{\em{All}}\/ 
the PG3 SRVs represent a group of stars
with a distribution similar, but  
intrinsically fainter, to the Miras.
A 1\mag0 shift towards brighter magnitudes
is an empirical patch for the luminosity difference
between SRVs and Miras. With this shift the foreground 
contamination for the SRVs is slightly smaller, but
probably the difference is not significant anymore.
In this case the PG3 SRVs are also contaminated with
foreground stars. 
\par
In Sect.~\ref{foreground} we argue that the latter possibility 
is preferred, but that a fraction of the foreground contamination
might be associated with the galactic bar.

\subsection{Periods and Amplitudes}
\label{PeriodAmplitudes}
Figure \ref{fig3} gives the period distributions of the various 
samples, used for comparison with the PG3 SRVs. 
The PG3 SRVs have periods comparable with
the short period tail of the PG3 Miras.
Their periods further overlap with 
the `red' field SRVs distinguished by KH92\mbox{\muspc\&\muspc}KH94,
but the PG3 SRVs have however a longer mean period.
\hfill\break
There is no significant difference between the period
distribution of the field and PG3 Miras, except for
a deficiency of PG3 variables with periods larger than 
300~days and is due to the 
spectral window for which the \BJ\ plates (Wess87) were taken,
see Sect.~\ref{PG3}. 
\par
The mean photo-visual amplitude, estimated from Plaut (\cite{Plaut71}),
for the PG3 SRVs is about 1\mag1. 
This is comparable to the V-amplitude
of the field SRVs in the same period range, but much smaller than the
amplitudes for the field Miras (5\mag5).
\par

\subsection{Period -- K Relation}
\label{PKrelation}
Figure \ref{fig4} shows the apparent K magnitude versus log\,P
diagram (hereafter referred to as PK$_0$-relation). The PG3 SRVs obey
the same PK$_0$-relation as the PG3 Miras
(Schultheis et~al.\ \cite{Schultheis96ea}). 
This figure suggests a common origin for the two samples.
Note that the straight line in Fig.~\ref{fig4}
is not a fit to the data! 
It shows the PK$_0$-relation for Sgr~I (Eq.~5, 
Glass et~al.\ \cite{Glass95ea}),
transformed from the SAAO to the ESO photometric system. 
We further adopted an extinction 
(\mbox{A$_V$\muspc=\muspc1\mag71} instead of
\mbox{A$_V$\muspc=\muspc1\mag87}) for the Sgr~I field.
The resulting PK$_0$-relation for Sgr~I in the 
ESO photometric system is: ${\rm K}_0\!=\!-3.47\log P+15.54$.
The dashed lines above and below the Sgr~I relation ($\pm\,1^m$)
are a combination of the expected scatter due to the depth of the bulge
\mbox{(${}^{+0.5}_{-0.6}$\ {\it mag})} 
and the dispersion in the magnitudes (not averaged, 
\mbox{$\pm$\muspc0\mag6}) 
of the PG3 variables.
The stars above the dotted line are foreground stars,
but see Sect.~\ref{foreground} for additional comments.
A few stars lie under the period-luminosity relation.
Ng\mbox{\muspc\&\muspc}Schultheis (\cite{NS97}) 
argue that those stars are 
located at the edge of the Sagittarius dwarf galaxy.
In the further analysis only those stars,
which are located between the two dashed lines, are considered.
\par

\subsection{Period - Colour Relation}
\label{PCrelation}
\begin{figure}
\vbox{\null\vskip16.0cm
\includegraphics{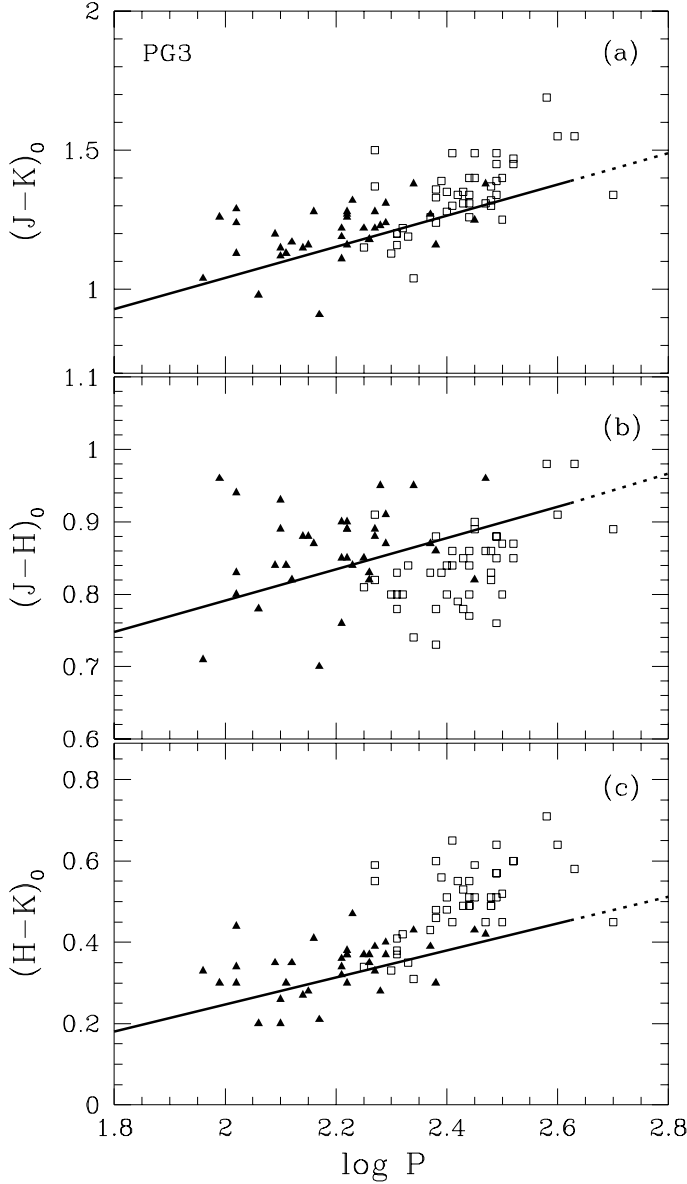}
\vfill}
\caption{Colour vs log P relations for the PG3 SRVs
\& Miras
for respectively {\bf a}) (J--H)$_0$, {\bf b}) (J--H)$_0$, and 
{\bf c}) (H--K)$_0$; only bulge stars are considered and
the symbols are the same as those used in Fig.~\ref{fig4}.
The thick straight lines are the LMC relations 
(Glass et~al.\ \cite{Glass95ea}) 
transformed from the SAAO(Mk3) to the ESO system 
(see Sect.~\ref{transformation}).
The dotted extension indicate an extrapolation of this relation
for \mbox{P\muspc$>$\muspc420$^d$}
}
\label{fig5}
\end{figure}

In Figs. \ref{fig5}a\to{c} the period-colour (PC) relations 
for the PG3 SRVs and Miras are shown for
(J--K)$_0$, (J--H)$_0$, and \mbox{(H--K)$_0$}, respectively.
The thick straight lines indicate
the LMC relation due to Feast et~al.\ (\cite{Feast89ea}) 
and Glass et~al.\ (\cite{Glass95ea}). 
In Fig.~\ref{fig5}a all the stars are slightly offset above the 
\mbox{P/(J--K)$_0$}  relation. 
In Fig.~\ref{fig5}b the Miras are below the P/(J--H)$_0$ relation,
while the SRVs are located slightly above. 
The (J--H)$_0$ colour for both the SRVs and Miras
appears to be independent of the period. 
In Fig.~\ref{fig5}c the PG3 SRVs appear to follow the LMC relation,
while the PG3 Miras are offset above. 
It is also possible that a fraction of the
SRVs follows the PG3 Mira P/(H--K)$_0$ relation, which is
steeper than the LMC relation.
An other fraction of SRVs lies clearly above such a relation.\par
\begin{figure}
\vbox{\null\vskip6.5cm
\includegraphics{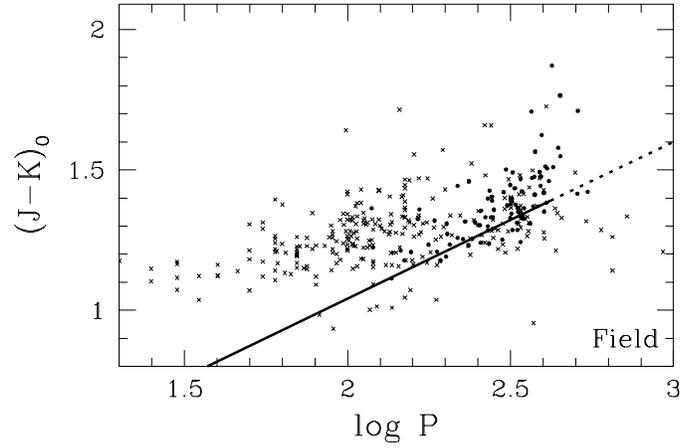}
}
\caption{(J--K)$_0$ vs log P relation for the field SRVs
(crosses; KH94 and Kerschbaum \cite{Kerschbaum95}) and field Miras 
(filled circles; Catchpole et~al.\ \cite{Catchpole79ea}). 
The thick straight line is the LMC relation due to 
Glass et~al.\ (\cite{Glass95ea}).
The photometry of the field Miras and the LMC relation
are transformed respectively from the SAAO(Mk1) and SAAO(Mk3)  
to the ESO system (see Sect.~\ref{transformation})
}
\label{fig6}
\end{figure}
For the PG3 Miras the mean offset from the LMC PC-relation
is \mbox{$\sim$\muspc0\mag05}. 
Within the transformation uncertainty of the LMC relation this
is comparable to the \mbox{$\sim$\muspc0\mag07} offset obtained by Bl92.
\hfill\break
Figure \ref{fig6} shows the P/(J--K)$_0$ relation for the field SRVs and the
field Miras.
The field Miras also follow the LMC relation, although
there is a slight offset of \mbox{$\sim$\muspc0\mag03} 
towards redder (J--K)$_0$.
This offset is not conclusive with regard to possible 
differences to the LMC or PG3 stars,
given the transformation uncertainties (field Mira and 
the LMC relation). 
\hfill\break
The 0\mag05 offset of the PG3 Miras translates with the theoretical 
period-colour relation from 
Wood et~al.\ (\cite{Wood91ea};
assuming comparable masses between the PG3 \& the field versus the LMC Miras)
in a mean metallicity of the PG3 and field Miras
$\sim$\muspc1.4 times as high as the LMC.
\par
\noindent
The majority of the field SRVs appear to  
follow a different PC-relation with a slope flatter than the field Miras.
But this might be an artifact, if the field SRVs are 
a non-homogeneous sample of fundamental 
mode pulsators with longer periods
and overtone pulsators with shorter periods. 
Since each mode has its own PC-relation, 
their combined distribution could well result in the flatter slope.
\par
\begin{figure}
\vbox{\null\vskip9.8cm
\includegraphics{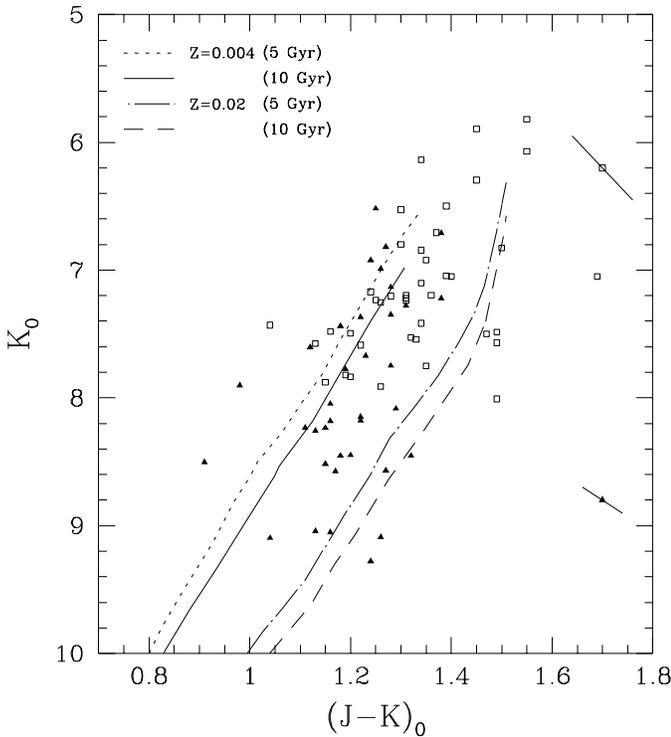}
}
\caption{K$_0$ vs (J--K)$_0$ Colour-Magnitude Diagram for the PG3 SRVs 
and Miras, the symbols are the same as those used in \mbox{Fig.~\ref{fig4}.}
The `error' bars attached to the two symbols on the right side indicate the 
effects of variability, the symbols do not correspond to the actual data.
Isochrones (Bertelli et~al.\ \cite{Bertelli94ea}), placed at 8~kpc distance
(Wesselink \cite{Wesselink87}, Reid \cite{Reid93}),
are shown for 5 and 10 Gyr populations
with \mbox{Z\muspc=\muspc0.004} and \mbox{Z\muspc=\muspc0.020}.
The near-infrared colours of those isochrones have been computed 
with an empirical $T_{\it eff}$-(J--K)$_0$ colour relation, 
see Ng et~al.\ (\cite{Ng98ea}; Table~1) for details about this relation.
}
\label{fig7}
\end{figure}

\subsection{Colour - Magnitude Diagram}
\label{CMD}
Figure \ref{fig7} shows the (K,J--K)$_0$\ CMD for the PG3 SRVs and Miras. 
Isochrones placed at 8~kpc distance
for 5 and 10~Gyr old stellar populations
with \mbox{Z\muspc=\muspc0.004} and \mbox{Z\muspc=\muspc0.020}
are displayed in this figure.
The isochrones from Bertelli et~al.\ (\cite{Bertelli94ea}) are used.
They converted their isochrones from the 
theoretical to the observational plane by convolving the 
near-infrared bands, as provided by 
Bessell\mbox{\muspc\&\muspc}Brett (\cite{Bessell88}),
with the spectral energy distributions from 
\mbox{Kurucz} (\cite{Kurucz92}) for 
temperatures higher then 4000~K. At lower temperatures they used observed 
spectra as described in Sect.~4 of 
Bertelli et~al.\ (\cite{Bertelli94ea})
and they combined the effective temperature scale from 
Ridgway et~al.\ (\cite{Ridgway80ea})
for the late M giants with the 
Lan\c{c}on\mbox{\muspc\&\muspc}Rocca-Volmerange (\cite{LRV92}) 
scale for the early M giants.
The lack of very red standards limits
the near-infrared colour transformations 
(Bressan\mbox{\muspc\&\muspc}Nasi \cite{BressanNasi95}) and
causes the colours of the \mbox{Z\muspc=\muspc0.02} isochrones
to `saturate' around \mbox{(J--K)$_0$\muspc$\simeq$\muspc1\mag35}. 
We derived a new, empirical 
$T_{\it eff}$-\mbox{(J--K)$_0$} colour relation 
by making a conservative fit through the $T_{\it eff}$ and 
\mbox{(J--K)$_0$} data
available for cool giants (see Ng et~al.\ 
\cite{Ng98ea} for details). This relation was adopted 
to compute the near infrared colours of the isochrones shown in 
Fig.~\ref{fig7}.

\par
The SRVs and Miras follow the trend 
indicated by the isochrones. 
SRVs and Miras with similar age and metallicity,
distributed around isochrones with comparable
age and metallicity, belong to the same population.
Note that variability moves the stars in an almost diagonal direction 
in the CMD. 
The upper limits for the
variation of the J--K colour around the light cycle 
is about \mbox{$\sim$\muspc0\mag20} for a Mira and
\mbox{$\sim$\muspc0\mag10} for a SRV 
(Hron\mbox{\muspc\&\muspc}Kerschbaum \cite{HK94}).
For the SRVs the amplitudes are too small to 
explain the scatter, while for the Miras the scatter 
might be for a large fraction due to their variability. 
\hfill\break
The uncertainties in the interstellar reddening is according to
Wess87 in the worst case 0\mag17 
in B$_{\rm J}$, which translates
in \mbox{$\sim$\muspc0\mag02} in K 
and \mbox{$\sim$\muspc0\mag03} in \mbox{J--K}.
Interstellar reddening therefore cannot explain the observed 
spread of \mbox{$\sim$\muspc0\mag20} in \mbox{J--K}. 
This spread on the other hand might be related to the intrinsic width of the 
instability strip. The intrinsic spread of the LMC PC relation
provides an upper limit. 
According to Kanbur et~al.\ (\cite{Kanbur97ea}) this spread
amounts for oxygen rich stars to 0\mag11 in \mbox{J--K}.
This is again smaller than the observed colour spread. 
\par
\begin{figure}
\vbox{\null\vskip8.8cm
\includegraphics{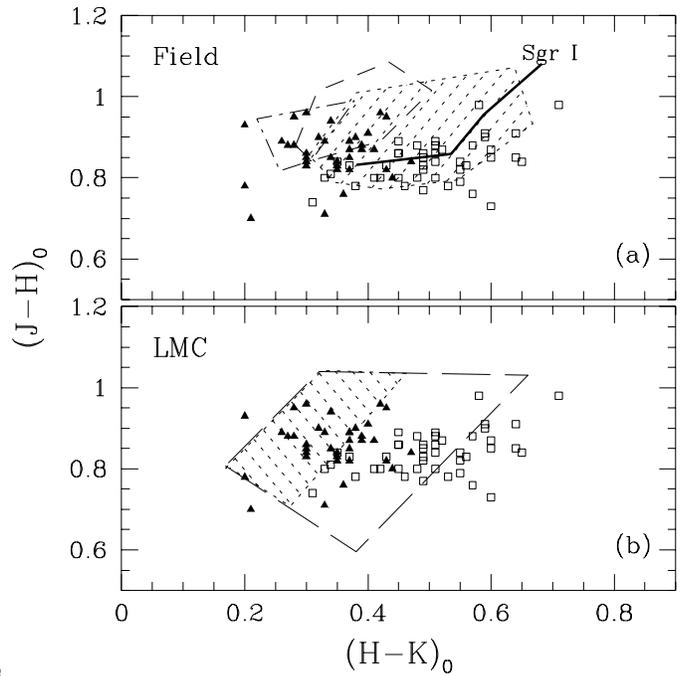}
}
\caption{(J--H)$_0$ vs (H--K)$_0$ Colour-Colour Diagram for the PG3 SRVs 
and Miras, the symbols are the same as those used in 
\mbox{Fig.~\ref{fig4}.} The 
polygon boxes show the location of the SRVs 
the Miras in different galactic environments:
({\bf a}) Field and
({\bf b}) LMC.
In frame {\bf a} the solid line
represents the location of the Sgr I Miras by 
Glass et~al.\ (\cite{Glass95ea}),
the dot-shaded box the Miras (Feast et~al.\ \cite{Feast89ea}), 
the dot-dashed box the `blue SRVs and the long-dashed box the
`red' SRVs (KH92\,\&\,KH94). In frame {\bf b} the dot-shaded box represents
the approximate location of the SRVs 
and the solid box the Miras from the 
Reid et~al.\ (\cite{Reid95ea}) LPVs (see Sect.~\ref{comparison})}
\label{fig8}
\end{figure}
Independently the variability of the SRVs, the uncertainties in the
interstellar reddening and the intrinsic width of the instability strip
cannot account for the observed colour spread. In combination even
an upper limit for the colour spread, which amounts to 0\mag15,
is not sufficient to explain the observed scatter.  
\par
The isochrones show that the effect of a 5~Gyr age difference
results merely in a shift of \mbox{$\sim$\muspc0\mag04} 
in the (J--K)$_0$ colour. On the other hand, 
metallicity differences result in larger shifts 
in the (J--K)$_0$ colour, e.g. \mbox{$\sim$\muspc0\mag20}
in Fig.~\ref{fig7}.
One might argue that the depth of the bulge would invalidate 
an analysis with isochrones, but one can verify easily that a
similar result is obtained by removing the differential distance effects.
For example, through an absolute calibration of the magnitude from 
the periods of the variables with the PK$_0$-relation at the distance
of the galactic centre, i.e. 8~kpc (Wesselink 1987, Reid 1993).
The reason for the similarity of the result is that it depends strongly
on the colour range covered and less on the magnitude (apparent or absolute)
of the stars. 
\par
The presence of a large spread in 
the metallicity could explain the distribution of the stars in the CMD. 
Due to the large spread in metallicity it 
is not possible to get a reliable age estimate.
The red edge is due to stars around solar metallicity,
while the stars at the blue edge have \mbox{Z\muspc=\muspc0.004}. 

\subsection{Colour -- Colour Diagram}
\label{CCD}
The colour-colour diagram in Fig.~\ref{fig8} demonstrates the difference
between the Miras and SRVs in PG3.
For comparison the different locations of
the Sgr~I Miras (Glass et~al.\ \cite{Glass95ea};
note that we adopted an extinction in agreement with 
\mbox{R$_0$\muspc=\muspc8~kpc}, see also Sect.~\ref{PKrelation}), 
the LMC LPVs
(Reid et~al.\ \cite{Reid95ea}) and the field Miras and SRVs 
(Feast et~al.\ \cite{Feast89ea}, KH92\,\&\,KH94) are indicated.
However, the shape of the LMC box is not well defined due
to the small number of stars used present in the region
\mbox{(J--H)$_0\!<$\muspc0\mag8}.
The PG3 SRVs are shifted with respect to the PG3 Miras to
bluer \mbox{(H--K)$_0$} and slightly redder (J--H)$_0$. 
\par
From the large similarity in period and amplitude between field `red' 
SRVs and PG3 SRVs one might expect that the PG3 SRVs 
will be located in the region
of the `red' field SRVs. Although there is some 
overlap, the PG3 SRVs appear to be 
on average bluer in both
colours than the `red' field SRVs. The PG3 SRVs extend to redder
colours than the LMC SRVs
\par
The PG3 Miras are more similar to the Sgr~I Miras than to the
comparison samples of field and LMC Miras. The PG3 stars do not extend to 
\mbox{(J--H)$_0$} colours as red as the Sgr~I Miras.
This could be related with deficiency of PG3 Miras with periods 
longer than 320 days. Within the uncertainties in the adopted colour
transformations PG3 and Sgr~I are comparable to each other.
\hfill\break
For \mbox{(J--H)$_0$\muspc$\la$\muspc0\mag75} and
\mbox{(H--K)$_0$\muspc$\la$\muspc0\mag50} 
the LMC is compared to the field and PG3 
abundant with relatively blue LPVs/Miras.
In addition, the LMC and the field have in contrast to PG3 and Sgr~I
for \mbox{(J--H)$_0$\muspc$\ga$\muspc0\mag90} and
\mbox{(H--K)$_0$\muspc$\la$\muspc0\mag60} 
in this region a significant number 
of LPVs/Miras.
In the following section we will argue 
that this is due to a combination of age and metallicity 
of the stars.
\par

\section{Discussion}
\label{Discussion}

\subsection{Miras: PG3 versus field, Sgr~I and LMC}
\label{Miras}
In Fig.~\ref{fig8} the PG3 Miras resemble
more the Sgr~I Miras than the field Miras and LMC LPVs. 
This is apparently in contradiction with 
Glass et~al.\ (\cite{Glass95ea},
Fig.~4c) who did not find any significant offset of the
Sgr~I Miras from  the LMC P/(J--K)$_0$ relation.
Note however, that the extinction correction is actually the
origin of this discrepancy.
Using $A_V$\muspc=\muspc1\mag71
the Sgr~I Miras will be 0\mag03 redder in \mbox{(J--K)$_0$}.
The estimated offset is now \mbox{$\sim$0\mag04} and 
within the uncertainties comparable to PG3.
\par
From Figs.~\ref{fig5}a and \ref{fig6} it is not clear 
if there is a significant offset
from the field Miras with respect to either the LMC or PG3,
while there is an offset between PG3 and the LMC. 
Figure~\ref{fig8} however shows that the field and PG3 Miras are not comparable
and that there are noticeable differences between all four groups of Miras: 
(i) in contrast to PG3 and Sgr~I, the field and
LMC Miras populate the region with \mbox{(J--H)$_0$\muspc$\ga$\muspc0\mag90},
(ii) the PG3 Miras extend to redder \mbox{(H--K)$_0$} colours than
all the other groups, and (iii) only the LMC Miras reach  
\mbox{(J--H)$_0$\muspc$\la$\muspc0\mag75}.

The blue \mbox{(J--H)$_0$} limit of the LMC Mira box is only defined 
by a few objects. 
The last point above is probably not a real difference but
induced by statistical fluctuations. Note that 
unidentified, hot carbon stars in the Reid et~al.\ (\cite{Reid95ea})
sample would be located in this area of the colour-colour diagram (see for 
example Fig.~5 from 
Costa\mbox{\muspc\&\muspc}Frogel \cite{CostaFrogel96}). 
However, the Reid et~al.\ stars 
are LPV's
while the blue carbon stars are not known to be large amplitude variables and
nothing is known about the variability of the blue carbon stars. This
point can be clarified only by more observational data. 

Statistical fluctuations can hardly be responsible for the other two
differences between the four groups of Miras. The redder \mbox{(H--K)$_0$} 
colours of the PG3 Miras could be due to metallicities higher than solar
but there is no evidence for this from the colour magnitude diagram. 
The majority of the field stars with solar metallicity have ages
between 1~Gyr and 8~Gyr (see Figs.~3\muspc\&\muspc4 from 
Ng\muspc\&{\muspc}Bertelli \cite{NB98}), 
while for PG3 our age estimates range
from 5~Gyr to 10~Gyr (Sect.~\ref{CMD}, Ng et~al.\ \cite{Ng95ea}). 
Thus the redder colour could be due to
an older age of the PG3 stars with \mbox{Z\muspc$\simeq$\muspc0.02}.

The lack of PG3 stars with \mbox{(J--H)$_0$\muspc$\ga$\muspc0\mag9} and 
\mbox{(H--K)$_0$\muspc$\la$\muspc0\mag5} cannot be 
accounted for by metallicity
effects alone, because this region is populated in the field as well as
in the LMC. Age differences might again be the reason. In the LMC the 
last major star formation occurred 6\to8~Gyr ago in some regions, 
while in other it happened only 2\to3~Gyr ago 
(Vallenari et~al.\ \cite{Vallenari96aea}b).
For field stars with \mbox{Z\muspc=\muspc0.008} the age ranges
from 2\to10~Gyr (see again Figs.~3\muspc\&\muspc4 from 
Ng\muspc\&{\muspc}Bertelli \cite{NB98}). 
Thus the lack of PG3 stars could be
due to a lack of stars with a metallicity in the range 
\mbox{Z\muspc$\simeq$\muspc0.008} and an age between \mbox{2\to5~Gyr}.

Although age is an attractive parameter to explain the differences
between the various groups of Miras we want to
emphasize that confirmation through a comparison with isochrones in 
the colour-colour diagram is still needed. This requires a proper 
calibration of the colours for the isochrones,
possibly combined with an improved description of the AGB-phase
(Ng et~al.\, \cite{Ng98ea}). 

All together 
the data are compatible with a metallicity range spanning 
from a quarter solar to approximately solar for the field and PG3 stars. 
The majority of the field stars
with metallicities around solar may be considerably younger
than their counterparts in PG3.
This would explain the 
smaller colour offset of the field Miras from the LMC
\mbox{P/(J--K)$_0$} relation in comparison to the colour offset for
the PG3 Miras. In this respect the apparent bluer
\mbox{(H--K)$_0$} colour in Fig.~\ref{fig8} of Sgr~I versus PG3
might be an indication for a slightly younger age for
the Miras in the Sgr~I area.

\subsection{SRVs: PG3 versus field and LMC}
\label{SRVs}
The PG3 SRVs are similar to `red' field SRVs in their periods and amplitudes. 
There are however marked differences: the slope of the
PK$_0$ and P/\mbox{(J--K)$_0$} relations for the PG3 SRVs is similar to that
of the PG3 and LMC Miras, while this does not appear to be the case for the
field SRVs (see KH92 and Fig.~\ref{fig4}). 
In addition, the colours of 
the PG3 SRVs are slightly bluer than those of the `red' field SRVs. 
The most plausible explanation for the colour differences is a higher
temperature of the PG3 SRVs compared to the field SRVs. 
A temperature difference could
explain the different behaviour in the P/\mbox{(J--K)$_0$} relations,
as outlined in Sect.~\ref{SRVvsMiras}.

The longer mean period of the PG3 stars is probably due to the 
larger homogeneity
of this sample relative to the field stars. This homogeneity concerns both the
variability classification and the pulsation mode (see below).

The result that the PG3 SRVs extend to redder colours than the LMC SRVs is
due to their higher average metallicity.

\subsection{SRVs versus Miras}
\label{SRVvsMiras}
Most field SRVs with pulsation periods 
below $\sim$\muspc200$^{\rm d}$ are cooler and 
partly brighter than field Miras with the same period
(see Fig.~\ref{fig6} and KH92).
This favours fundamental mode
pulsation for the Miras and overtone pulsation for the SRVs. 
Similar evidence
for variables in the LMC was presented by 
Wood\mbox{\muspc\&\muspc}Sebo (\cite{WS96}, hereafter WS96).
Their results indicate that the fundamental mode pulsation 
is consistent with stellar masses $\la\!2 M_\odot$.
This might be in contradiction with the results obtained by
van~Leeuwen et~al.\ (\cite{vLeeuwen97ea}). Their analysis indicate that
the majority of the Miras are first overtone pulsators.
However, their sample did not include SRVs and 
is furthermore biased to stars with large radii.
For stars in fundamental mode with shorter periods (like the PG3 SRVs)
the data for the smaller radii are lacking, due to observational limitations.
Therefore, the results from van~Leeuwen et~al.\ are not necessarily
in disagreement with WS96.
\par
An interpretation of the PG3 stars in terms of pulsation modes has to be 
in agreement with the behaviour of the SRVs and Miras in the PK$_0$ and
\mbox{P/(J--K)$_{0}$} diagrams.
The PG3 SRVs would be brighter and redder than the Miras at a given period, 
if the PG3 Miras are fundamental mode pulsators and the PG3 SRVs are  
overtone pulsators. If Miras and SRVs pulsate in the same mode,
a systematically lower metallicity of the SRVs would increase 
their temperature and make them bluer, but at the same time
their periods would be smaller at a given luminosity. This
would introduce a luminosity difference relative to the Miras at a 
given period but would only shift
the SRVs along their P/\mbox{(J--K)$_0$} relation. 
\hfill\break
Therefore, our result that the PG3 SRVs are an extension of the 
PG3 Mira PK$_0$ and P/(J--K)$_0$ relations,
can only be explained by adopting the same metallicity range and pulsation 
mode for
the Miras and SRVs. In view of the similarities between field and PG3
Miras, fundamental mode pulsation is more plausible for the PG3 stars.
\hfill\break
In addition, we conclude that the PG3 SRVs are {\em{not}}\/ 
the analogs of the field SRVs (see Figs.~\ref{fig5}\,\&\,\ref{fig6}). 
\par

\subsection{\em{The metallicity of PG3 Miras and SRVs}}
\label{metallicityPG3}
From star counts and metal-rich globular cluster studies
(Ng \cite{Ng94}; Bertelli et~al.\ \cite{Bertelli95ea}, 
\cite{Bertelli96ea}; Minniti \cite{Minniti95}) 
one expects a gradient of metal-rich stars 
towards the galactic centre. 
BW is located closer to the galactic centre
and has a larger number of high metallicity stars 
with respect to PG3 (Ng et~al.\ \cite{Ng96aea}, \cite{Ng97ea}).
The period-colour relation indicates that the mean metallicity 
of the PG3 variables is about 1.4 times larger than the LMC 
mean metallicity.
A comparison of both the PG3 and Sgr~I (Glass et~al.\ \cite{Glass95ea})
period-(J--K)$_0$ relation relative to the LMC relation 
shows that a small trend might be present.
But the uncertainties in
the extinction correction for Sgr~I are   
larger than in PG3. An uncertainty in the extinction of 
\mbox{$\sim$\muspc0\mag1\to0\mag2} 
in one filter would give an error in the colour, i.e.~(J--K)$_0$,
of a few hundredth of a magnitude.
Together with the uncertainties in the transformation,
the remaining difference between Sgr~I and PG3 
period-colour relation is not significant to conclude
from the present data,
that there is a metallicity gradient towards the galactic centre.
\par
The PG3 SRVs and Miras extend to redder (H--K)$_0$
than the LMC LPVs. This
indicates that the mean metallicity for the PG3 LPVs is slightly larger
than for the LMC which is also consistent with Fig.~\ref{fig7}.
The bulk of PG3 Miras cover the same position
as the Sgr~I Miras by Glass et~al.\ (\cite{Glass95ea}). 
They should have a similar 
age and metallicity as the Sgr~I Miras, which presumably 
have solar-type metallicity.
This implies that the whole metallicity range 
from intermediate to solar is possible
in PG3 and the LMC,
but as mentioned in Sect.~\ref{Miras} there are perhaps differences in age 
between PG3 and LMC stars of comparable metallicity.

\subsection{Hints from galactic structure}
\label{galstruct}
The field SRVs have a scale height of 230 pc (KH92). This is slightly
smaller than the scale height of 
260\muspc$\pm$\muspc30~pc for the PG3 Miras with a detection in IRAS
(Bl92). These values are comparable with the 250~pc obtained 
by Habing (\cite{Habing88}) for AGB stars and they are consistent with the
scale height for disc giants.
The whole PG3 Mira sample, on the other hand, has a scale height 
of 300\muspc$\pm$\muspc50~pc.
Although the previous values are within the uncertainties one
has to consider, that differentially there is a noticeable
difference between the Miras with and without an IRAS detection.
With the ages and scale heights, determined for the 
stellar populations in the disc 
(Ng \cite{Ng94}; Ng et~al.\ \cite{Ng95ea}, \cite{Ng96aea}, \cite{Ng97ea}),
the age of the field SRVs and the PG3 Miras with an detection in
IRAS is between \mbox{4.5\to7.0~Gyr}
with a metallicity ranging from \mbox{Z\muspc=\muspc0.008\to0.015}. 
The age range for the whole PG3 Mira sample is estimated 
\mbox{4.5\to7.5~Gyr}. 
These ages appears to be in agreement with an age considerably less than 
10~Gyr, estimated by 
Harmon\mbox{\muspc\&\muspc}Gilmore (\cite{HarmonGilmore88}) 
for the bulge IRAS sources with initial 
masses larger than 1.3~$\cal M_\odot$.
But Whitelock et~al.\ (\cite{Whitelock91ea}) showed, 
that their lower limit estimate for the initial
mass is more likely an upper limit for the majority of the stars. 
This gives a lower age limit 
\mbox{$\sim$\muspc4~Gyr} (Bertelli et~al.\ \cite{Bertelli94ea}), 
which is consistent with the age deduced from the scale heights.
\par
The very large metallicity spread causes that age estimates
are very susceptible. Even ages as old as 
16~Gyr estimated by Bl92 are possible. 
With such an old age the bluest SRVs and Miras
should be very metal-poor.
This would lead to a contradiction that long period variables cannot be
present, because they are not found in old, metal-poor globular clusters.
Therefore, the PG3 Miras and SRVs cannot be old and very
metal-poor, but this does not rule out the intermediate and 
solar metallicity cases. 
The period for the PG3 Miras ranges from 180\to320 days, which is comparable
with the periods of the Miras found in metal-rich globular clusters
(Feast\mbox{\muspc\&\muspc}Whitelock \cite{FW87}, 
Whitelock et~al.\ \cite{Whitelock91ea}).
The \mbox{8\to9~Gyr} age of these clusters 
(Ng et~al. \cite{Ng96cea}d)
gives an initial mass of about 1~$\cal M_\odot$, which is
consistent with the upper limit obtained from the scale height for
these stars.
\par
On the other hand, if the disc density towards the 
galactic centre is lower than expected from
a double exponential density profile
(Bertelli et~al.\ \cite{Bertelli95ea}; 
Kiraga et~al.\ \cite{Kiraga97ea};
Ng \cite{Ng94}\mbox{\muspc\&\muspc}\cite{Ng97a}; 
Ng et~al.\ \cite{Ng95ea}, \cite{Ng96aea}; 
Paczy\'nski\mbox{\muspc\&\muspc}Udalski \cite{PaczynskiUdalski97};
Paczy\'nski et~al.\ \cite{Paczynski94ea}), this could imply that the 
PG3 SRVs and Miras are not due to a disc population.
As argued above they are also not related to a very
old, metal-poor population. With an upper age
of about 8~Gyr, the PG3 Miras are in that case likely related with
the `bar' population identified by Ng et~al.\ (\cite{Ng96aea}b).
The age and metallicity spread for this population 
(t\muspc=\muspc8\to9 Gyr; \mbox{Z\muspc=\muspc0.005\to0.03})
might imply that the variables in PG3 are located 
in the outer regions of the `bar'.
A complication is that one should be
careful with the semantics related with the `bar'.
If the stars of this population can be found 
in PG3, i.e. $\sim$\muspc1.5~kpc out of the 
galactic plane, this population cannot
be originating from a bar as found in bar-like galaxies.
Strictly spoken, this should be referred to as a triaxial structure.
\par
The bluest SRVs and Miras might have a metallicity of 
\mbox{Z\muspc$\simeq$\muspc0.005} with an age $\sim$\muspc9~Gyr.
Although the uncertainty in the ages is between 1\to3~Gyr,
the large metallicity spread seems to indicate that all ages 
between \mbox{5\to10~Gyr} are possible in the metallicity
range \mbox{Z\muspc=\muspc0.005\to0.03}.
The great similarity in Fig.~\ref{fig8}b between the distributions 
of the variables from PG3 and the LMC suggests a 
comparable age and metallicity for both samples.
Vallenari et~al.\ (\cite{Vallenari96aea}b) find indications for 
enhancements of the star formation rate in the LMC 
at ages as old as \mbox{6\to8~Gyr}, but in other regions
the bulk of star formation has occurred only 
\mbox{2\to3~Gyr} ago.
\par
This raises the question if stars younger than 5~Gyr are
present in the galactic bulge/bar. 
The bluer \mbox{(J--H)$_0$} colour of the Sgr~I Miras with respect to those
from PG3 might indicate a younger age for the former, 
but as pointed out in Sect.~\ref{CCD} the colour difference is 
probably due to uncertainties in the colour transformations.
\hfill\break
The presence of carbon stars could be an indication for
a young age, because
the work from Marigo et~al.\ (\cite{Marigo96aea}b 
and references cited therein) 
indicates that carbon stars cannot be much older than 4~Gyr.
The carbon stars (L199,S283) identified in our sample of 
LPVs and those identified by Azzopardi et~al.\ (\cite{Azzopardi91ea})
might therefore be an indication for the presence of stars 
younger than 5~Gyr in the bulge/bar.
Ng\muspc\&\muspc{Schultheis} (\cite{NS97})
argue that S283 is actually related to the Sagittarius dwarf 
galaxy found by Ibata et~al.\ (\cite{IGI94}).
Possibly the same argument holds for PG3 variable L199.
Furthermore it was argued that
the sample of carbon stars from Azzopardi et~al.\ (\cite{Azzopardi91ea})
is probably not associated with the bulge.
In the bulge they would be bolometrically 
\mbox{$\sim$\muspc2\mag5} too faint, 
but associated with the Sagittarius dwarf galaxy
their luminosities are comparable to carbon stars 
found in other dwarf galaxies.
\par
If all the bulge carbon stars are related to 
the Sagittarius dwarf galaxy (Ng \cite{Ng97b}, \cite{Ng98}) 
this would imply that they
are absent in the bulge.
In contrast to the field and LMC sample 
this implies the absence of a major star formation 
epoch less than 4~Gyr ago.

\subsection{PG3 and the foreground stars}
\label{foreground}
As already described in Sect.~\ref{magdistribution} the 
foreground contamination of PG3 Miras
is either significantly larger or comparable with the PG3 SRVs. 
Figure~\ref{fig7} hints that both groups belong to the same population
and a similar fraction of foreground stars is therefore expected.
This implies that the SRVs are one magnitude fainter 
than the Miras. The K magnitude distribution of the two groups
are in this case comparable. The asymmetry suggests that
more stars are found nearby. This could imply 
a high foreground contamination as suggested by BL92,
but most likely it is due to stars located in the
nearby side of the triaxial structure mentioned above.
Some of the nearby stars in the sample
belongs therefore to the same population 
as the bulge SRVs and Miras and
should not have been treated as foreground contamination. 

\section{Summary}
\label{Summary}
$\circ$\ We have shown that PG3 SRVs are not the analogs to the field SRVs.
The comparison of the \mbox{P/(J--K)$_0$} relation
of the two SRV groups shows that they do not
obey the same \mbox{P/(J--K)$_0$} relation. 
In addition their location in a colour-colour diagram
differs slightly. All together this indicates 
a different nature between the two SRV groups. 
\hfill\break
$\circ$\ The PG3 SRVs form a short period extension to the 
Miras PK$_0$ and PC-relations.
This indicates that the PG3 Miras and SRVs are 
both pulsating in the same mode, possibly the fundamental.
\hfill\break
$\circ$\ The field SRVs (the `blue' and the majority of the `red' group) 
are likely overtone pulsators.
\hfill\break
$\circ$\  The metallicities of the PG3 SRVs and Miras
span the range from intermediate to approximately solar.
\hfill\break
$\circ$\ The age possibly covers a range from \mbox{4\to10~Gyr}.
From the absence of LPVs in metal-poor globular clusters
it is argued that the PG3 SRVs and Miras in the bulge 
are likely not older than 10~Gyr. 
From the upper mass limit of the bulge IRAS sources 
and the possible absence of bulge carbon stars
one obtains a lower age limit of 4~Gyr.
\hfill\break
$\circ$\ Field and PG3 Miras follow
the same P/(J--K)$_0$ relation and cover the same region in the
(J--H)$_0$ vs (H--K)$_0$ diagram. Therefore, the metallicity
of the field and PG3 Miras should overlap each other. 
The Miras and SRVs in PG3 follow the Sgr~I PK$_0$-relation. 
This confirms independently the work of 
Whitelock et~al.\ (\cite{Whitelock91ea}) 
and Glass et~al.\ (\cite{Glass95ea}):
they found no difference in the PL-relation for different
galactic environments.
\par
The following question arises: are there SRVs in PG3, 
similar to those found in the disc?
The presence or absence of these stars might provide an
indication of the age of the stars in PG3.
The missing SRVs might be hidden among the irregular variables. 
According to Wess87 
those are variable stars with little or no trace of
periodicity for which the amplitudes do not exceed 1\mag5. 
For verification, a detailed study of these stars is desired.
\par
Another question which arises concerns the nature of the large spread 
in metallicity. The large range of ages seems to indicate that 
both young and old stars can be present with intermediate up to 
solar metallicity. 
Even the presence of more massive, young
stars, which can be metal-poorer than older stars, is
possible. In a closed box model one expects an increasing metallicity 
towards younger ages. Is this an indication that
a closed box model is not applicable? 
What is the origin of this behaviour?

\begin{acknowledgements}
The authors thank H.J.~Habing for his 
encouragements of this collaboration
and the referees (Drs. Feast and Whitelock)
for constructive suggestions. 
P.R.~Wood is acknowledged for comments on the pulsation modes of AGB
variables.
The research of J.~Hron, F.~Kerschbaum 
and M.~Schultheis is supported by the
Austrian Science Fund projects 
P9638--AST and S7308.
Y.K.~Ng thanks the
Institut f\"ur \mbox{Astronomie} der Universit\"at Wien
and the department of Astronomy from the University of Padova,
where part of this research was carried out, for their hospitality.
The university of Wien provided financial support for Ng's
research visit to Vienna and 
ANTARES, an astrophysics network funded by the HCM programme of the
European Community, supported Ng's research visits to Padova. 
At the IAP-CNRS and at the Padova Astronomical Observatory 
the research of Ng was supported 
by respectively HCM grant CHRX-CT94-0627
and TMR grant ERBFMRX-CT96-0086
from the EC.
\end{acknowledgements}

\hyphenation{Slijk-huis}

\vfill
\onecolumn
%
%
\addtolength{\vsize}{1.0cm}
\noindent
{\bf{Table 2:}} Near-infrared photometry for
the stars in field \#3 of the Palomar-Groningen Variable Star
Survey (Plaut 1971). Column 1 lists the stellar identifier, 
adopted from Wesselink (1987);
column 2 gives the identification made by Plaut (1971);
column 3--7 gives the JHKL$^\prime$M photometry,
typical errors are $\sim$ 0\mag02 in JHK, $\sim$ 0\mag1 in L$^\prime$
and $>$ 0\mag2 in M;
column 8 gives the observing run identifier (see table 1);
column 9 gives the period determined by Wesselink (1987)
if available; and column 10 gives the quality flag related 
with the period and the identification of the star 
(Q=0: no doubt about the determined period and classification,
Q=1: classification is correct but alternative period is possible,
Q=2: period determination is correct but the classification is doubtful,
Q=3: both period determination and classification are unreliable)
\begin{center}
\begin{tabular}{|rlll|rlrlrlrlrl|clrlrl|}
\hline
\multicolumn{2}{|c}{Name}
&\multicolumn{1}{c}{PL71}&
&\multicolumn{2}{c}{J}
&\multicolumn{2}{c}{H}
&\multicolumn{2}{c}{K}
&\multicolumn{2}{c}{L$^\prime$}
&\multicolumn{2}{c|}{M}
&\multicolumn{1}{c}{Obs.}&
&\multicolumn{2}{c}{P}
&\multicolumn{1}{c}{Q}
&\multicolumn{1}{l|}{}
\\
\hline
\hline
S6&
&SRa&
&9.53&
&8.63&
&8.31&
&7.7&
& &
&4&
&128.70&
&0&
\\
&
& &
&9.61&
&8.68&
&8.33&
&7.8&
&8.9&
&5&
& &
& &
\\
S40&
&SRa&
&9.51&
&8.56&
&8.21&
&7.8&
&  &
&1&
&137.01&
&3&
\\
&
& &
&9.57&
&8.61&
&8.30&
&8.2&
&6.6&
&3&
& &
& &
\\
 &
& &
&9.53&
&8.63&
&8.27&
&7.9&
& &
&4&
& &
& &
\\
S46&
&SRa&
&9.23&
&8.31&
&7.72&
&6.8&
&6.1&
&5&
& &
&3&
\\
S59&
&L&
&9.55&
&8.71&
&8.31&
&7.8&
&7.5&
&4&
&163.69&
&0&
\\
S70&
&SRa&
&8.35&
&7.41&
&7.02&
&6.5&
&6.5&
&4&
&166.52&
&0&
\\
&
& &
&8.57&
&7.55&
&7.11&
&6.6&
&6.3&
&5&
& &
& &
\\
S96&
&SRb&
&8.69&
&7.83&
&7.36&
&6.6&
&6.6&
&5&
& &
&3&
\\
S99&
&SRa&
&10.32&
&9.53&
&9.16&
&8.9&
&6.7&
&4&
&92.16&
&0&
\\
S144&
&SRa&
&9.71&
&8.72&
&8.41&
&8.0&
&8.3&
&4&
&165.75&
&2&
\\
 &
& &
&9.61&
&8.62&
&8.36&
&8.0&
&7.1&
&5&
& &
& &
\\
S147&
&SRa &
&9.44&
&8.47&
&8.13&
&8.1&
& &
&1&
&165.75&
&0&
\\
 &
& &
&8.93&
&8.18&
&8.00&
&8.0&
& &
&3&
& &
& &
\\
 &
& &
&9.41&
&8.47&
&8.12&
&7.7&
&6.5&
&4&
& &
& &
\\
S197&
&L&
&9.45&
&8.45&
&8.10&
&7.6&
&9.4&
&5&
& &
&3&
\\
S283&
&SRa&
&10.64&
&10.12&
&9.85&
&9.8&
&7.5&
&4&
&137.04&
&0&
\\
&
& &
&11.69&
&11.06&
&10.93&
& &
& &
&5&
& &
& &
\\
S289&
&SRb&
&10.34&
&9.43&
&9.10&
&8.5&
&6.9&
&4&
&104.78&
&0&
\\
S325&
&SRa&
&9.45&
&8.57&
&8.24&
&7.9&
&7.8&
&5&
&240.10&
&0&
\\
S326&
&C&
&10.97&
&10.13&
&9.79&
& &
& &
&4&
&86.87&
&2&
\\
S328&
&SRa&
&9.14&
&8.22&
&7.84&
&7.6&
&6.9&
&5&
&161.30&
&0&
\\
S398&
&SRa&
&9.52&
&8.60&
&8.20&
&7.6&
&6.9&
&4&
&178.57&
&0&
\\
S458&
&SRa&
&9.34&
&8.46&
&8.21&
&7.9&
&7.9&
&4&
& &
&3&
\\
S463&
&SRa&
&9.30&
&8.48&
&8.02&
&7.4&
&6.6&
&4&
&115.70&
&3&
\\
S472&
&SRa&
&11.30&
&10.33&
&10.18&
&9.3&
& &
&5&
&210.30&
&0&
\\
S510&
&SRa&
&9.08&
&8.22&
&7.97&
&7.9&
&7.0&
&4&
&114.89&
&0&
\\
S514&
&SRa&
&9.84&
&8.92&
&8.52&
&8.2&
&8.3&
&4&
&123.27&
&0&
\\
S538&
&SRa&
&9.80&
&8.86&
&8.57&
&8.6&
& &
&5&
&125.50&
&0&
\\
S539&
&SRa&
&8.33&
&7.27&
&6.83&
&6.2&
&5.9&
&4&
& &
&3&
\\
S561&
&SRa&
&8.79&
&7.90&
&7.50&
&7.0&
&6.4&
&4&
&184.00&
&0&
\\
S568&
&SRa&
&9.93&
&9.03&
&8.64&
&7.8&
&6.6&
&4&
&132.55&
&0&
\\
S588&
&SRa&
&8.20&
&7.12&
&6.68&
&6.1&
&5.8&
&4&
&178.00&
&2&
\\
S601&
&SRa&
&9.83&
&8.92&
&8.52&
&8.1&
&7.4&
&4&
&181.52&
&0&
\\
S639&
&SRa&
&8.76&
&7.81&
&7.40&
&6.8&
&6.5&
&5&
&167.30&
&0&
\\
S662&
&M&
&9.92&
&9.01&
&8.51&
&7.7&
&7.9&
&4&
&169.91&
&0&
\\
S680&
&SRb&
&9.86&
&9.17&
&8.96&
&8.3&
& &
&5&
& &
&3&
\\
S714&
&SRb&
&8.33&
&7.36&
&6.99&
&6.7&
&6.8&
&5&
& &
&3&
\\
S719&
&SRa&
&7.89&
&7.02&
&6.56&
&6.0&
&6.0&
&4&
&279.77&
&0&
\\
S728&
&SRa&
&7.26&
&6.32&
&5.57&
&4.6&
&4.3&
&1&
& &
&3 &
\\
&
& &
&7.21&
&6.34&
&5.71&
&4.6&
&4.4&
&3&
& &
& &
\\
&
& &
&7.15&
&6.17&
&5.49&
&4.5&
&4.6&
&5&
& &
& &
\\
S791&
&L&
&8.77&
&7.75&
&7.28&
&6.8&
&7.3&
&4&
&219.89&
&0&
\\
S861&
&SRb&
&10.16&
&9.25&
&8.89&
&8.6&
&9.2&
&5&
& &
&3&
\\
S910&
&SRa&
&7.67&
&6.71&
&6.27&
&5.7&
&5.6&
&5&
&224.30&
&0&
\\
S915&
&SRa&
&8.40&
&7.39&
&6.98&
&6.6&
&6.8&
&5&
& &
&3&
\\
S930&
&SRa&
&9.37&
&8.52&
&8.19&
&7.7&
&9.1&
&5&
&149.60&
&1&
\\
S932&
&SRa&
&9.95&
&8.98&
&8.68&
&8.9&
& &
&5&
& &
&3&
\\
\hline
\end{tabular}
\end{center}
\eject
\addtolength{\vsize}{1.0cm}
\noindent
{\bf{Table 2.} {\it continued ...}}\\
\begin{center}
\begin{tabular}{|rlll|rlrlrlrlrl|clrlrl|}
\hline
\multicolumn{2}{|c}{Name}
&\multicolumn{1}{c}{PL71}&
&\multicolumn{2}{c}{J}
&\multicolumn{2}{c}{H}
&\multicolumn{2}{c}{K}
&\multicolumn{2}{c}{L$^\prime$}
&\multicolumn{2}{c|}{M}
&\multicolumn{1}{c}{Obs.}&
&\multicolumn{2}{c}{P}
&\multicolumn{1}{c}{Q}
&\multicolumn{1}{l|}{}
\\
\hline
\hline
S942&
&M&
&8.18&
&7.28&
&6.78&
&6.2&
&6.5&
&5&
&176.00&
&1&
\\
S964&
&SRb&
&8.69&
&7.83&
&7.36&
&6.6&
&6.6&
&5&
& &
&3&
\\
&
& &
&8.69&
&7.82&
&7.32&
&6.6&
&6.3&
&5&
& &
& &
\\
S967&
&L&
&10.11&
&9.18&
&8.89&
&8.4&
&7.1&
&5&
&335.90&
&0&
\\
S969&
&SRa&
&7.03&
&6.06&
&5.62&
&5.2&
&5.4&
&2&
& &
&3 &
\\
 &
& &
&7.01&
&6.06&
&5.65&
&5.2&
&5.5&
&3&
& &
& &
\\
 &
& &
&7.15&
&6.18&
&5.70&
&5.3&
&5.3&
&5&
& &
& &
\\
S972&
&SRa&
&10.65&
&9.80&
&9.32&
&9.3&
&7.5&
&5&
&105.10&
&0&
\\
S984&
&SRa&
&8.07&
&7.07&
&6.57&
&5.9&
&5.9&
&5&
& &
&3&
\\
S988&
&SRa&
&8.05&
&7.11&
&6.75&
&6.4&
&6.6&
&5&
& &
&3&
\\
S1002&
&SRa&
&8.31&
&7.38&
&6.98&
&6.6&
&6.0&
&4&
&194.23&
&0&
\\
S1005&
&L&
&7.69&
&6.69&
&6.28&
&5.9&
&5.8&
&5&
&199.60&
&0&
\\
S1008 &
&SRa&
&8.13&
&7.21&
&6.76&
&6.4&
&6.6&
&1&
&232.14&
&0&
\\
 &
& &
&8.20&
&7.28&
&6.86&
&6.4&
&6.7&
&3&
& &
& &
\\
S1012&
&SRa&
&8.32&
&7.39&
&7.03&
&6.5&
&6.9&
&5&
& &
&3&
\\
S1016 &
&SRa &
&5.68&
&4.69&
&4.25&
&3.8&
&3.9&
&2&
& &
&3&
\\
 &
& &
&5.72&
&4.71&
&4.28&
&3.8&
&3.9&
&3&
& &
& &
\\
 &
& &
&5.84&
&4.79&
&4.32&
&3.8&
&4.0&
&5&
& &
& &
\\
S1059&
&SRa&
&9.12&
&8.21&
&7.78&
&7.5&
&6.4&
&5&
&144.10&
&0&
\\
S1101&
&SRa&
&10.50&
&9.52&
&9.18&
&9.1&
& &
&4&
&79.70&
&1&
\\
S1128 &
&SRa &
&9.77&
&8.88&
&8.59&
&8.4&
& &
&3&
&97.00&
&0&
\\
 &
& &
&11.29&
&10.51&
&10.30&
& &
& &
&4&
& &
& &
\\
S1176&
&SRa&
&8.54&
&7.60&
&7.18&
&6.7&
&6.5&
&5&
&184.10&
&0&
\\
S1181&
&SRa&
&9.04&
&8.08&
&7.74&
&7.3&
&6.5&
&5&
& &
&3&
\\
S1203&
&SRa&
&11.79&
&10.98&
&10.81&
&9.7&
&7.6&
&4&
&134.92&
&0&
\\
S1204 &
&SRa &
&8.73&
&7.78&
&7.34&
&6.8&
&6.1&
&2&
&197.00&
&0&
\\
 &
& &
&8.68&
&7.73&
&7.31&
&6.7&
&6.9&
&3&
& &
& &
\\
S1266&
&SRa&
&9.74&
&9.19&
&9.06&
&8.3&
&6.7&
&4&
&146.89&
&0&
\\
&
& &
&9.23&
&8.31&
&7.99&
&7.6&
&6.8&
&5&
& &
& &
\\
S1293&
&SRa&
&10.32&
&9.39&
&9.09&
&9.1&
& &
&4&
&141.27&
&0&
\\
S1295&
&SRa&
&9.56&
&8.60&
&8.30&
&7.9&
&7.1&
&5&
&131.00&
&1&
\\
S1358&
&SRa&
&9.92&
&8.99&
&8.60&
&8.1&
&7.2&
&4&
&167.00&
&0&
\\
S1410&
&SRa&
&8.56&
&7.57&
&7.14&
&6.6&
&6.3&
&4&
&202.65&
&2&
\\
S1470&
&SRa&
&8.61&
&7.69&
&7.36&
&7.0&
&6.3&
&4&
&184.08&
&0&
\\
&
& &
&8.65&
&7.75&
&7.41&
&7.0&
&7.2&
&5&
& &
& &
\\
S1489&
&&
&8.89&
&7.91&
&7.52&
&7.2&
&8.6&
&4&
&74.65&
&0&
\\
S1517&
&&
&9.05&
&8.04&
&7.72&
&7.1&
&6.2&
&5&
&188.80&
&0&
\\
S1523&
&&
&8.53&
&7.28&
&6.86&
&6.4&
&6.5&
&4&
&223.32&
&0&
\\
S1528&
&&
&9.29&
&8.32&
&7.97&
&8.0&
&6.4&
&5&
& &
&3&
\\
S1555&
&&
&8.63&
&7.56&
&7.10&
&6.6&
&7.5&
&4&
&149.15&
&3&
\\
&
&&
&8.76&
&7.71&
&7.22&
&6.7&
&7.7&
&5&
& &
& &
\\
S1579&
&&
&11.53&
&10.84&
&10.67&
& &
& &
&4&
& &
&3&
\\
S1638&
&&
&7.03&
&6.02&
&5.62&
&5.2&
&5.7&
&5&
& &
&3&
\\
S1644&
&&
&9.50&
&8.50&
&8.13&
&7.6&
& &
&5&
&104.20&
&0&
\\
S1709&
&&
&9.89&
&8.88&
&8.59&
&8.2&
&8.2&
&5&
& &
&3&
\\
S1788&
&&
&10.60&
&9.47&
&9.18&
&9.2&
&7.7&
&5&
& &
&3&
\\
S1855&
&&
&6.69&
&5.64&
&5.23&
&4.8&
&5.1&
&5&
&227.60&
&1&
\\
S1869&
&&
&8.27&
&7.23&
&6.77&
&6.2&
&6.3&
&4&
&298.17&
&0&
\\
S1907&
&&
&9.61&
&8.62&
&8.26&
&8.0&
&6.6&
&5&
&163.80&
&0&
\\
S1933&
&&
&9.49&
&8.51&
&8.20&
&8.1&
&7.0&
&5&
& &
&3&
\\
S1966&
&&
&10.52&
&9.48&
&9.15&
&8.5&
& &
&4&
&97.55&
&0&
\\
S1991&
&&
&8.94&
&7.92&
&7.68&
&7.4&
&8.7&
&5&
&124.70&
&0&
\\
S2082&
&&
&8.30&
&7.29&
&6.900&
&6.6&
&6.6&
&5&
& &
&3&
\\
S2096&
&&
&11.54&
&10.79&
&10.54&
& &
& &
&4&
&472.72&
&0&
\\
\hline
\end{tabular}
\end{center}

\end{document}